\documentclass[10pt]{article}  

\usepackage[dvips]{graphics,graphicx,color}

\usepackage{epsf}

\begin{document}

\title{Quantum Monte Carlo study of the Ne atom and the Ne$^+$ ion}

\author{N.~D.~Drummond, P.~L\'opez~R\'{\i}os, A.~Ma, J.~R.~Trail, \\
G.~Spink, M.~D.~Towler, and R.~J.~Needs \\ Theory of Condensed Matter
Group, Cavendish Laboratory, University of Cambridge, \\ J.~J.~Thomson
Avenue, Cambridge CB3 0HE, United Kingdom}

\maketitle

\begin{abstract}
We report all-electron and pseudopotential calculations of the
ground-state energies of the neutral Ne atom and the Ne$^+$ ion using
the variational and diffusion quantum Monte Carlo (VMC and DMC)
methods.  We investigate different levels of Slater-Jastrow trial wave
function: (i) using Hartree-Fock orbitals, (ii) using orbitals
optimized within a Monte Carlo procedure in the presence of a Jastrow
factor, and (iii) including backflow correlations in the wave
function.  Small reductions in the total energy are obtained by
optimizing the orbitals, while more significant reductions are
obtained by incorporating backflow correlations.  We study the
finite-time-step and fixed-node biases in the DMC energy, and show
that there is a strong tendency for these errors to cancel when the
first ionization potential (IP) is calculated.  DMC gives highly
accurate values for the IP of Ne at all the levels of trial wave
function that we have considered.
\end{abstract}

\newpage

\section{Introduction}

Accurate approximations to the many-electron wave function are
required as inputs for the variational and diffusion quantum Monte
Carlo (VMC and DMC) electronic-structure methods \cite{foulkes_2001}.
The quality of these ``trial'' wave functions determines both the
statistical efficiency of the methods and the final accuracy that can
be obtained, and one of the main technical challenges in the field of
quantum Monte Carlo (QMC) simulation is to develop more accurate trial
wave functions.

We report a detailed VMC and DMC study of the ground states of the Ne
atom and the Ne$^+$ ion, in which several different forms of trial
wave function have been used, and both all-electron and
pseudopotential calculations have been performed.  The difference
between the ground-state energies of the neutral atom and the positive
ion gives the first ionization potential (IP), which is known
accurately from experiments \cite{kaufman_1972}.  The IP's calculated
using different wave functions are compared with the essentially exact
nonrelativistic IP \cite{chakravorty_1993} and the experimental
value. We also demonstrate the degree to which the DMC fixed-node and
time-step errors cancel when the IP is calculated.

The basic form of wave function that we study consists of a product of
Slater determinants for spin-up and spin-down electrons containing
Hartree-Fock (HF) orbitals, multiplied by a positive Jastrow
correlation factor.  We consider the effects of optimizing the
orbitals in the presence of the Jastrow factor
\cite{umrigar_1988,filippi_2000}, and the effects of including
backflow correlations in the wave function
\cite{feynman_1954,feynman_1956,kwon_1993}.  We have also carried out
some tests using multideterminant wave functions
\cite{filippi_1996,schautz_2002}, but we were not able to lower the
DMC energy using this approach.

The rest of this paper is organized as follows.  In Sec.~\ref{sec:qmc}
we briefly review the VMC and DMC methods, while in
Sec.~\ref{sec:trial_wave_functions} we describe the different levels
of Slater-Jastrow wave function that we investigate. In
Sec.~\ref{sec:neon_dt_bias} we study time-step bias in DMC
calculations of the total energy and IP of Ne.  The total energies and
IP's calculated using different trial wave functions are compared in
Sec.~\ref{sec:ne_neplus_energies}.  Finally, we draw our conclusions
in Sec.~\ref{sec:conclusions}.

We use Hartree atomic units, $\hbar=|e|=m_e=4\pi \epsilon_0=1$,
throughout.  All of our QMC calculations were performed using the
\textsc{casino} code \cite{casino}.

\section{QMC methods \label{sec:qmc}}

The VMC energy is calculated as the expectation value of the
Hamiltonian operator with respect to a trial wave function, the
integrals being performed by a Monte Carlo method.  In DMC, the
imaginary-time Schr\"odinger equation is used to evolve a set of
configurations towards the ground-state distribution.  Fermionic
symmetry is maintained by the fixed-node approximation
\cite{anderson_1976}, in which the nodal surface of the wave function
is constrained to equal that of a trial wave function.  Our DMC
algorithm is essentially that of Umrigar \textit{et al.}\
\cite{umrigar_1993a}. Various modifications to the DMC Green's
function were proposed by Umrigar \textit{et al.}\ to reduce the
time-step errors in all-electron calculations, and we investigate the
effects of these modifications in Sec.~\ref{sec:neon_dt_bias}.

For most of our pseudopotential calculations we used a nonrelativistic
HF Ne pseudopotential (effective core potential), although for our
investigations of time-step errors we used a Dirac-Fock
averaged-relativistic effective potential (AREP)
\cite{trail_2005a,trail_2005b}. These pseudopotentials are finite at
the origin, making them particularly suitable for use in QMC
calculations.  There is evidence \cite{greeff_1998} that neglecting
correlation effects entirely when constructing pseudopotentials, as in
HF and Dirac-Fock theory, provides better pseudopotentials for use in
correlated valence calculations than density-functional-theory-derived
pseudopotentials. Part of the motivation for performing the
calculations reported here is to test the Ne pseudopotentials in
correlated valence calculations.  The contributions to the total
energy from the nonlocal components of the pseudopotential within DMC
were calculated using the locality approximation \cite{hurley_1987},
which leads to errors that are second order in the error in the trial
wave function \cite{mitas}.

\section{Trial wave functions \label{sec:trial_wave_functions}}

\subsection{Basic Slater-Jastrow wave function with HF orbitals
\label{subsec:slater-jastrow}}

The Slater-Jastrow wave function may be written as
\begin{equation}
\label{eq:slater-jastrow}
\Psi ({\bf R})=\exp[J({\bf R})] \sum_i \mu_i D^\uparrow_i({\bf R})
D^\downarrow_i({\bf R}),
\end{equation}
where ${\bf R} \equiv \left( {{\bf r}_1,\ldots ,{\bf r}_N} \right)$
denotes the spatial coordinates of all the electrons, $\exp[J( {\bf R}
)]$ is the Jastrow correlation factor, $D^\uparrow_i({\bf R})$ and
$D^\downarrow_i({\bf R})$ are Slater determinants of orbitals for
spin-up and spin-down electrons, and the $\{ \mu_i \}$ are expansion
coefficients.  The Jastrow factor, which describes the dynamic
correlation of the electrons, is an explicit function of the
interparticle distances.  We use the form of Jastrow factor described
in Ref.\ \cite{drummond_2004}. In our calculations the Slater
determinants $\{ D^\sigma_i \}$ were formed from single-particle
orbitals obtained from restricted open-shell HF calculations using
numerical integration on radial grids.  The all-electron HF
calculations were performed using the code of Fischer
\cite{gaigalas_1996} and the pseudopotential HF calculations were
performed using our own code.  The variable parameters in the trial
wave functions were determined by minimizing the unreweighted variance
of the energy \cite{umrigar_1988,kent_1999,drummond_2005}.

We also performed tests using multideterminant trial wave functions.
The determinants were obtained from multiconfiguration HF calculations
performed by numerical integration on radial grids, the expansion
coefficients [$\{ \mu_i \}$ in Eq.~(\ref{eq:slater-jastrow})] being
re-optimized in the presence of the Jastrow factor using unreweighted
variance minimization \cite{drummond_2005}. During the optimization
process, we constrained the coefficients of the determinants within
each configuration state function to have the same magnitude, so that
the symmetry of the wave function was maintained.  We were able to
achieve significantly lower VMC energies than we could with the
single-determinant Slater-Jastrow wave function, but the DMC energies
were higher than the single-determinant ones.  All numerical results
reported in this article were obtained with a single term in the
determinant expansion.

\subsection{Modification of the HF orbitals
\label{subsec:orbital_optimization}}

Altering parameters in the orbitals moves the nodal surface of the
trial wave function and therefore changes the value of the fixed-node
DMC energy.  This is not the case for parameters in the Jastrow
factor.  Optimizing the orbitals in the presence of a Jastrow factor
generally improves the nodal surface of the wave function and
therefore lowers the DMC energy, which satisfies a variational
principle with respect to errors in the nodal surface
\cite{reynolds_1982,foulkes_1999}. Unfortunately, optimizing
parameters that affect the nodal surface by unreweighted variance
minimization is problematic.  At each iteration, a minimization is
performed using a fixed set of sampling points in the configuration
space, and the nodal surface deforms continuously as the parameter
values are altered.  In the course of this deformation, the nodal
surface may pass through one of the fixed configurations, at which
point the estimate of the unreweighted variance of the local energy
($\Psi^{-1} \hat{H} \Psi$, where $\hat{H}$ is the Hamiltonian
operator) diverges.  This makes it difficult to locate the true
minimum of the unreweighted variance.  It is possible to improve the
situation by using the weight-limiting scheme described in
Ref.~\cite{drummond_2005}, which effectively removes configurations
whose local energies are diverging from the set used to calculate the
variance.  Alternatively, the reweighted variance-minimization
algorithm can be used, where each configuration is weighted by the
ratio of the square of the current wave function to the probability
density according to which the configurations were distributed, so
that the calculated variance is an estimate of the actual variance of
the energy. Reweighted variance minimization does not suffer from the
problem of divergences in the variance but, nevertheless, the
unreweighted algorithm gives significantly lower variational energies
in practice, and we used this method for all the optimizations
reported here.  However, the difficulty in locating the global minimum
of the variance with respect to parameters that affect the nodal
surface makes it more difficult to optimize the orbitals for two
different systems to the same level of accuracy.

The atomic orbitals that we have used can be expressed in the form
\begin{equation}
\label{eq:orbital}
\psi_{nlm}({\bf r}) = \left[ \rho^{\rm HF}_{nl}(r) + \Delta
\rho_{nl}(r) \right] r^l Y_{lm}(\theta,\phi),
\end{equation}
where we have used spherical polar coordinates, with $r$, $\theta$,
and $\phi$ being the radial, polar, and azimuthal coordinates of the
point ${\bf r}$.  The origin is chosen to lie at the nucleus.
$Y_{lm}(\theta,\phi)$ is the $(l,m)$th spherical harmonic, and
$\rho^{\rm HF}_{nl}(r)$ is the HF radial function.  (Note that, in
practice, we use appropriate linear combinations of spherical
harmonics with a given $l$ in order to construct real orbitals.)  The
modification to the HF radial function is written as
\begin{equation}
\Delta \rho_{nl}(r) = \left( \sum_{j=0}^N c_{jnl} \, r^j \right) \exp
\left( \frac{-A_{nl}r^2}{1+B_{nl}r} \right),
\label{eqn:orbmod_fn_defn}
\end{equation}
where $A_{nl}$, $B_{nl}$ and $\{ c_{0nl},\ldots,c_{Nnl} \}$ are
optimizable parameters.  Obviously we must have $A_{nl}>0$ and $B_{nl}
\geq 0$, otherwise the orbital will be non-normalizable.  The chosen
form of the modification to the radial function has considerable
variational freedom and decays exponentially at large distances.  The
HF orbitals minimize the energy in the absence of a Jastrow factor,
although they are no longer optimal when a Jastrow factor is included;
however we expect $\Delta \rho_{nl}(r)$ to remain small in this case.
After some experimentation we decided to use $N=3$ for the
all-electron calculations and $N=6$ for the pseudopotential ones, as
these choices allowed us to achieve the lowest variational energies in
our tests.

Kato derived a condition on the spherical average of the wave function
about a bare nucleus of atomic number $Z$ at the origin
\cite{kato_1957}:
\begin{equation}
\left( \frac{\partial \langle\Psi\rangle}{\partial r_{i}}
\right)_{r_{i}=0} = -Z \langle\Psi\rangle_{r_{i}=0},
\label{eq:kato}
\end{equation}
where $r_i=|{\bf r}_i|$ and $\langle\Psi\rangle$ is the spherical
average of the many-body wave function about $r_i=0$.  If the wave
function is nonzero at the nucleus (as is the case for the
ground-state wave functions studied in the present work), then it is
possible to carry out accurate and stable QMC calculations if and only
if the Kato cusp condition [Eq.~(\ref{eq:kato})] is satisfied
\cite{foulkes_2001}. It is easy to show that a Slater-Jastrow wave
function obeys the electron-nucleus Kato cusp condition if each
orbital satisfies Eq.~(\ref{eq:kato}) individually and the Jastrow
factor is cuspless at the nucleus.  (It is preferable to impose the
electron-nucleus cusp condition via the orbitals rather than the
Jastrow factor \cite{drummond_2004}.) The spherical averages of
orbitals with $l \neq 0$ are zero, and therefore these orbitals obey
the Kato cusp condition automatically, but we must impose
Eq.~(\ref{eq:kato}) on the radial parts of the $s$ orbitals.  This
gives us the requirement that $c_{1n0}=-Zc_{0n0}$, where $Z$ is the
atomic number of the atom in an all-electron calculation and $Z=0$ for
a calculation using a pseudopotential which is finite at the nucleus.

\subsection{Backflow correlations \label{subsec:backflow}}

We have also investigated the effect of incorporating backflow
correlations in the trial wave function.  Classical backflow is
related to the flow of a fluid around a large impurity, and its
quantum analog was first discussed by Feynman \cite{feynman_1954} and
Feynman and Cohen \cite{feynman_1956} in the contexts of excitations
in $^{4}$He and the effective mass of a $^{3}$He impurity in liquid
$^{4}$He.  Backflow correlations have previously been used in trial
wave functions for fermion QMC simulations of two-dimensional
\cite{kwon_1993} and three-dimensional \cite{kwon_1998} electron
gases, and metallic hydrogen \cite{pierleoni_2004}. The original
derivation of backflow correlations by Feynman \cite{feynman_1954} was
based on the idea of local current conservation, although they can
also be derived from an imaginary-time evolution argument
\cite{kwon_1993,holzmann_2003}.

The backflow correlations are described by replacing the electron
coordinates $\{ {\bf r}_i \}$ in the Slater determinants of
Eq.~(\ref{eq:slater-jastrow}) by ``quasiparticle'' coordinates $\{
{\bf x}_i \}$, defined to be
\begin{equation}
{\bf x}_i = {\bf r}_i + \sum_{j\neq i} \left[ F(r_{ij},r_i,r_j)({\bf
r}_i-{\bf r}_j) + G(r_{ij},r_i,r_j) {\bf r}_j \right],
\end{equation}
where $F$ and $G$ are functions that contain parameters to be
determined by variance minimization, and $r_{ij} = |{\bf r}_i -{\bf
r}_j|$.  In our work, $F(r_{ij},r_i,r_j)$ and $G(r_{ij},r_i,r_j)$
consist of smoothly truncated polynomial expansions in $r_i$, $r_j$,
and $r_{ij}$, of the same form as the isotropic terms that occur in
the Jastrow factor introduced in Ref.\ \cite{drummond_2004}.

In our all-electron calculations the electron-nucleus cusp conditions
are enforced via the orbitals in the Slater determinants, and in order
to preserve the cusp conditions when backflow is included we ensure
that $F$ and $G$ go smoothly to zero as an electron approaches the
nucleus.  We use pseudopotentials that are smooth at the nucleus and
we therefore choose $G$ to be smooth at the nucleus in our
pseudopotential calculations.  The electron-electron Kato cusp
conditions are enforced by the Jastrow factor in our calculations, and
the backflow functions are chosen to preserve these conditions.  Our
implementation of backflow correlations for systems containing many
atoms will be described elsewhere \cite{lopez_rios_2005}.

\section{Investigations of time-step errors \label{sec:neon_dt_bias}}

We have studied the time-step bias in all-electron and Dirac-Fock AREP
pseudopotential DMC calculations of the ground-state energies of the
Ne atom and the Ne$^+$ ion \cite{trail_2005a,trail_2005b}.  Figures
\ref{fig:ne_tstep_bias} and \ref{fig:ne_plus_tstep_bias} show the DMC
energy obtained at different time steps, both with and without the
modifications to the all-electron DMC Green's function proposed in
Ref.~\cite{umrigar_1993a}.  The all-electron energies enter the
small-time-step regime (where the bias in the DMC energy is nearly
linear in the time step) at time steps less than about $0.005$~a.u.
The root-mean-square distance diffused at each time step of
$0.005$~a.u.\ is about $0.12$~a.u., which is close to the Bohr radius
of Ne ($1/Z=0.1$\,a.u.); this is the length scale relevant to the core
($1s$) electrons, suggesting that they are responsible for much of the
time-step bias.  The time-step bias remains small up to much larger
time steps in the pseudopotential calculations, again implying that
the $1s$ electrons are responsible for most of the bias in the
all-electron results.  The contribution to the time-step bias from the
core region is expected to be large, because the energy scale of the
innermost electrons is greater than the energy scale of the outer
electrons, although this is offset by the fact that the wave function
is relatively accurate in the vicinity of the nucleus.

It is clear from Figs.~\ref{fig:ne_tstep_bias} and
\ref{fig:ne_plus_tstep_bias} that the modifications to the
all-electron DMC Green's function successfully eliminate much of the
time-step bias due to the innermost electrons at small time steps.  On
the other hand, the form of the time-step bias is almost the same for
the Ne atom and the Ne$^+$ ion, suggesting that the bias will cancel
out when the IP is calculated, irrespective of whether the Green's
function modifications are used.  It can be seen in
Fig.~\ref{fig:ne_ion_tstep_bias}, which shows the IP as a function of
time step, that this is indeed the case.  Since the bias is largely
caused by the innermost electrons, which are in almost identical
environments in Ne and Ne$^+$, this strong cancellation is to be
expected.

If a basic Slater-Jastrow wave function is used then the
zero-time-step limit of the DMC energy is determined by the nodes of
the HF wave function, independent of the Jastrow factor, whereas the
DMC energy at finite time steps does depend on the Jastrow factor.
However, provided that similar levels of Jastrow factor are used for
systems to be compared (with the core electrons in the same
configuration), it is reasonable to expect a large degree of
cancellation of bias to occur when energy differences are taken at
finite time steps.

\section{Energies of Ne and Ne$^+$ \label{sec:ne_neplus_energies}}

In Table 1 we present values for the total nonrelativistic energies of
Ne and Ne$^{+}$ and the nonrelativistic IP of Ne, calculated using a
number of different electronic-structure methods.  For the
all-electron atom and ion, we give results obtained using HF theory
(AHF), density-functional theory using the local spin-density
approximation (DFT-LSDA), Perdew-Wang (DFT-PW91), and
Becke-Lee-Yang-Parr (DFT-BLYP) exchange-correlation functionals,
all-electron VMC and DMC using a basic Slater-Jastrow wave function
with HF orbitals (AVMC and ADMC), all-electron VMC and DMC using a
Slater-Jastrow wave function with optimized orbitals (AOVMC and
AODMC), all-electron VMC and DMC with HF orbitals and backflow
correlations (ABVMC and ABDMC), and all-electron VMC and DMC with
optimized orbitals and backflow correlations (AOBVMC and
AOBDMC\@). For the pseudopotential calculations, which used a
nonrelativistic HF pseudopotential \cite{trail_2005a,trail_2005b}, we
give results obtained using HF theory (PHF), VMC and DMC using a basic
Slater-Jastrow wave function with HF orbitals (PVMC and PDMC), VMC and
DMC using a Slater-Jastrow wave function with optimized orbitals
(POVMC and PODMC), VMC and DMC with HF orbitals and backflow
correlations (PBVMC and PBDMC), and VMC and DMC with optimized
orbitals and backflow correlations (POBVMC and POBDMC\@).  For each of
the VMC calculations we give the total number of parameters in the
trial wave function that were optimized using unreweighted variance
minimization, along with the expectation value of the variance of the
energy.  Furthermore, we give the fraction of the correlation energy
retrieved by the wave function. In the all-electron calculations, the
correlation energy is defined to be the difference between the HF
energy and the exact nonrelativistic energy.  In the pseudo-Ne
calculations, the correlation energy is defined to be the difference
between the HF energy and the PBDMC energy.  The DFT-LSDA IP is taken
from Ref.~\cite{perdew_1992}, while the DFT-PW91 and DFT-BLYP results
are taken from Ref.~\cite{grabo_1992}.  The ``exact'' nonrelativistic
(NR) infinite-nuclear-mass energies are taken from
Ref.~\cite{chakravorty_1993}.

All the DMC energies quoted in Table 1 have been extrapolated to zero
time step.  We used a range of small time steps and performed linear
extrapolations of the energies to zero time step.

Optimizing the orbitals reduces the VMC energy, but does not have a
significant effect on the DMC energy, except in the case of the
pseudopotential calculation with backflow correlations, where
optimizing the orbitals actually increases the DMC energy, presumably
because the very large number of variable parameters adversely affects
the optimization.  The VMC variance is very slightly reduced by
optimizing the orbitals in the all-electron calculations, but not in
the pseudopotential calculations.  (Note that, as discussed in
Ref.~\cite{drummond_2005}, the unreweighted variance-minimization
algorithm does not generally minimize the expected variance of the
energy.)  The inclusion of backflow correlations reduces the errors in
the VMC energies by a factor of about 2, and reduces the variances of
the VMC energies by a factor of between 2 and 3.  For both
pseudopotential and all-electron calculations, the percentage of the
total correlation energy retrieved within VMC is slightly larger for
Ne than for Ne$^+$, although the difference is smaller in DMC\@.

The previous lowest VMC energy in the literature for the all-electron
Ne atom is due to Huang \textit{et al.}\ \cite{huang_1997}, who used a
single-determinant trial wave function with optimized orbitals and
Jastrow factor, corresponding to our AOVMC level of calculation, and
obtained an energy of $-128.9029(3)$~a.u., which amounts to 91.11(8)\%
of the correlation energy.  Our AOVMC energy of $-128.90334(9)$~a.u.\
(91.23(2)\% of the correlation energy) is very similar.  We obtain a
VMC energy of $-128.9205(2)$~a.u.\ (95.62(5)\% of the correlation
energy) at the AOBVMC level, so that backflow retrieves about
50.1(7)\% of the remaining correlation energy.  Note also that the VMC
energies obtained with backflow are very similar to the DMC energies
without backflow, for both the all-electron and pseudopotential
calculations.  This indicates that VMC with backflow may be a useful
level of theory, because VMC calculations are significantly less
costly than DMC ones, and VMC has advantages for calculating
expectation values of quantities other than the energy.

At the DMC level backflow retrieves about 38(2)\% of the correlation
energy missing at the AODMC level.  Our AOBDMC energy of
$-128.9290(2)$~a.u.\ (97.80(5)\% of the correlation energy) is
significantly lower than the previous lowest DMC energy in the
literature of $-128.9243(8)$~a.u.\ (96.6(2)\% of the correlation
energy) \cite{huang_1997}.

The increase in the complexity of the wave function resulting from the
inclusion of backflow correlations is apparent from the three- to
five-fold increase in the number of parameters in the wave function.
However, the use of backflow or orbital optimization does not result
in any clear improvement to the VMC or DMC IP's.  The all-electron DMC
IP's obtained with the different forms of wave function are very close
to one another.  The inclusion of backflow in the pseudopotential
calculations results in a small increase in the IP (away from the
experimental result).  This may reflect the fact that it is harder to
optimize a backflow function for Ne$^+$, where the symmetry between up
and down spins is broken, or it may be an indication of the error
inherent in the pseudopotential approximation.

Irrespective of the form of trial wave function used, or whether
all-electron or pseudopotential calculations are performed, or whether
the VMC or DMC method is used, all of the QMC calculations give
excellent values for the IP\@.  In the worst cases the errors in the
VMC nonrelativistic IP's are 0.43(4)\% (all-electron) and 0.60(3)\%
(pseudopotential) and the errors in the DMC nonrelativistic IP's are
0.28(4)\% (all-electron) and 0.35(4)\% (pseudopotential).  The
considerably larger errors in the HF IP (8\%) and DFT IP's (1.7\% for
the best case of the BLYP functional) demonstrate the importance of
describing electron correlation accurately when calculating the IP\@.

The correction to the IP of Ne from the finite mass of the nucleus is
of order $10^{-5}$~a.u., which is negligible on the scale of interest.
The total relativistic correction to the IP of Ne has been estimated
to be $-0.00196$~a.u.\ \cite{chakravorty_1993}, which is significant
on the scale of interest: see Fig.~\ref{fig:ne_ion_tstep_bias}.
Adding this correction to our best DMC data, obtained at the AOBDMC
level, gives an IP of $0.7945(2)$~a.u., which is only
$0.0020(2)$~a.u.\ ($0.054(5)$~eV) larger than the experimental value
of $0.792481$~a.u.\ ($21.5645$~eV) \cite{kaufman_1972}. Using the same
relativistic correction for the pseudopotential POBDMC data gives an
IP which is $0.0026(3)$~a.u.\ ($0.071(8)$~eV) larger than experiment.
The error in the IP due to the use of the pseudopotential is therefore
about $0.0006(4)$~a.u.\ ($0.016(11)$~eV), which is less than 0.1\% of
the IP.

\section{Conclusions \label{sec:conclusions}}

We have performed all-electron and pseudopotential VMC and DMC
calculations of the total energies of Ne and Ne$^+$ using basic
Slater-Jastrow wave functions with HF orbitals, Slater-Jastrow wave
functions with optimized orbitals, and wave functions including
backflow correlations.  We have found that optimizing the orbitals
makes a small improvement to the VMC energy, but has very little
effect on the DMC energy.  The HF orbitals give nearly optimal
single-determinant nodal surfaces for the ground states of Ne and
Ne$^+$, although this cannot be expected to hold for other systems.
On the other hand, including backflow correlations lowers both the VMC
and DMC energies significantly, giving us the lowest VMC and DMC
energies for all-electron Ne in the literature to date.  However, the
improvements in the total energies of Ne and Ne$^+$ hardly affect the
calculated IP, because they almost exactly cancel when the energy
difference is evaluated.  Overall, the calculated IP's are slightly
too large, because the wave functions for Ne$^+$ are not quite as
accurate as those for Ne.

Including backflow correlations for Ne and Ne$^+$ reduced the error in
the VMC total energy by a factor of about 2 and reduced the variance
of the VMC energy by a factor of between 2 and 3.  Reducing the
variance improves the statistical efficiency of VMC calculations,
while reducing the error in the VMC total energy improves the
statistical efficiency of DMC calculations
\cite{ceperley_1986,ma_2005}.  The incorporation of backflow is
expected to improve the QMC estimates of all expectation values, not
just the energy.  Including backflow correlations is costly, however,
because every element in the Slater determinant has to be recomputed
each time an electron is moved, whereas only a single column of the
Slater determinant has to be updated after each move when the basic
Slater-Jastrow wave function is used.  In the case of Ne, including
backflow correlations increased the computational cost per move in VMC
and DMC by a factor of between 4 and 7.  However, the reduction in the
variance meant that the number of moves required to obtain a fixed
error bar in the energy was smaller; hence, overall, including
backflow correlations increased the time taken to perform the
calculations by a factor of between 2 and 3.  Backflow functions
introduce additional variable parameters into the trial wave function,
making the optimization procedure more difficult and costly.  Nodal
surfaces obtained from HF or DFT orbitals do not suffer from
statistical errors.  By contrast, the parameters in the backflow
functions are optimized by a Monte Carlo method, and therefore the
nodal surface is subject to stochastic noise.  Furthermore, although
Jastrow factors are easy to optimize using unreweighted variance
minimization with a fixed distribution of configurations, parameters
that affect the nodal surface are relatively difficult to optimize,
because the local energy diverges at the nodes.

We have investigated the time-step bias in the DMC total energies of
Ne and Ne$^+$ and the IP of Ne.  The time-step errors in the
pseudopotential calculations were relatively small, indicating that
the bias in the all-electron total energy was mainly due to the $1s$
electrons.  In IP calculations using a basic Slater-Jastrow wave
function with HF orbitals the time-step bias largely canceled out when
the difference of energies was taken, irrespective of whether the
modifications to the all-electron DMC Green's function proposed in
Ref.~\cite{umrigar_1993a} were used.  The time-step errors in the IP
were considerably smaller in the pseudopotential calculations than the
all-electron ones.

Most importantly, our results demonstrate the superb accuracy of the
DMC method for this problem.  After including a correction for
relativistic effects, our all-electron IP of Ne is about
$0.0020$~a.u.\ ($0.054$~eV) larger than the experimental value
\cite{kaufman_1972} of $0.792481$~a.u.\ ($21.5645$~eV).  Our
pseudopotential IP is about $0.0026$~a.u.\ ($0.071$~eV) larger than
the experimental value, demonstrating the accuracy of the
pseudopotentials we use \cite{trail_2005a,trail_2005b}.

\section{Acknowledgments}

Financial support has been provided by the Engineering and Physical
Sciences Research Council of the United Kingdom.  N.D.D.\ acknowledges
financial support from Jesus College, Cambridge. P.L.R.\ gratefully
acknowledges the financial support provided through the European
Community's Human Potential Programme under contract
HPRN-CT-2002-00298, RTN ``Photon-Mediated Phenomena in Semiconductor
Nanostructures.''  Computer resources have been provided by the
Cambridge-Cranfield High Performance Computing Facility.

\newpage

\textbf{Table}

\begin{center}
\begin{footnotesize}
\begin{tabular}{lccr@{.}lr@{.}lr@{.}lr@{.}lr@{.}lr@{.}lr@{.}l}
\hline \hline

& \multicolumn{2}{c}{No.~param.} & \multicolumn{4}{c}{Energy (a.u.)} &
  \multicolumn{4}{c}{\%age corr.~en.} & \multicolumn{4}{c}{Variance
  (a.u.)} & \multicolumn{2}{c}{} \\

\raisebox{1ex}[0pt]{Meth.} & Ne & Ne$^+$ & \multicolumn{2}{c}{Ne} &
\multicolumn{2}{c}{Ne$^+$} & \multicolumn{2}{c}{Ne} &
\multicolumn{2}{c}{Ne$^+$} & \multicolumn{2}{c}{Ne} &
\multicolumn{2}{c}{Ne$^+$} & \multicolumn{2}{c}{\raisebox{1ex}[0pt]{IP
(a.u.)}} \\

\hline

Exact NR   & --  & --  & $-128$&$9376$     & $-128$&$1431$     &
$100$&$0$ & $100$&$0$ & \multicolumn{2}{c}{--}          &
\multicolumn{2}{c}{--}         & $0$&$7945$      \\

DFT-LSDA & --  & --  & \multicolumn{2}{c}{--} & \multicolumn{2}{c}{--}
& \multicolumn{2}{c}{--} & \multicolumn{2}{c}{--} &
\multicolumn{2}{c}{--} & \multicolumn{2}{c}{--} & $0$&$812$ \\

DFT-PW91 & --  & --  & \multicolumn{2}{c}{--} & \multicolumn{2}{c}{--}
& \multicolumn{2}{c}{--} & \multicolumn{2}{c}{--} &
\multicolumn{2}{c}{--} & \multicolumn{2}{c}{--} & $0$&$812$ \\

DFT-BLYP & --  & --  & \multicolumn{2}{c}{--} & \multicolumn{2}{c}{--}
& \multicolumn{2}{c}{--} & \multicolumn{2}{c}{--} &
\multicolumn{2}{c}{--} & \multicolumn{2}{c}{--} & $0$&$808$ \\

\hline

AHF       & --  & --  & $-128$&$54710$ & $-127$&$81781$ & $0$&$0$ &
$0$&$0$ & \multicolumn{2}{c}{--}          & \multicolumn{2}{c}{--} &
$0$&$72928$ \\

AVMC      & 75  & 79  & $-128$&$8983(2)$  & $-128$&$1016(1)$  &
$89$&$94(5)$ & $87$&$24(3)$ & $1$&$12(2)$   & $1$&$047(1)$ &
$0$&$7967(2)$   \\

AOVMC     & 88 & 92   & $-128$&$90334(9)$ & $-128$&$10760(6)$ &
$91$&$23(2)$ & $89$&$09(2)$ & $0$&$9740(9)$ & $0$&$938(3)$ &
$0$&$79574(11)$ \\

ABVMC     & 205 & 209 & $-128$&$9190(2)$  & $-128$&$1213(2)$  &
$95$&$24(5)$ & $93$&$30(6)$ & $0$&$421(1)$  & $0$&$406(2)$ &
$0$&$7979(3)$   \\

AOBVMC    & 218 & 222 & $-128$&$9205(2)$  & $-128$&$1240(2)$  &
$95$&$62(5)$ & $94$&$13(6)$ & $0$&$391(1)$  & $0$&$379(3)$ &
$0$&$7965(3)$   \\

ADMC      & --  & --  & $-128$&$9233(2)$  & $-128$&$1266(2)$  &
$96$&$34(5)$ & $94$&$93(6)$ & \multicolumn{2}{c}{--}          &
\multicolumn{2}{c}{--} & $0$&$7967(3)$    \\

AODMC     & --  & --  & $-128$&$9237(2)$  & $-128$&$1273(2)$  &
$96$&$44(5)$ & $95$&$14(6)$ & \multicolumn{2}{c}{--}          &
\multicolumn{2}{c}{--} & $0$&$7964(3)$   \\

ABDMC     & --  & --  & $-128$&$9287(2)$  & $-128$&$1321(1)$  &
$97$&$72(5)$ & $96$&$62(3)$ & \multicolumn{2}{c}{--}          &
\multicolumn{2}{c}{--} & $0$&$7966(2)$   \\

AOBDMC    & --  & --  & $-128$&$9290(2)$  & $-128$&$1325(1)$  &
$97$&$80(5)$ & $96$&$74(3)$ & \multicolumn{2}{c}{--}          &
\multicolumn{2}{c}{--} & $0$&$7965(2)$   \\

\hline

PHF       & --  & --  & $-34$&$59111$  & $-33$&$86129$  & $0$&$0$ &
$0$&$0$ & \multicolumn{2}{c}{--} & \multicolumn{2}{c}{--} &
$0$&$72982$  \\

PVMC      & 53  & 61  & $-34$&$88930(4)$  & $-34$&$09453(4)$  &
$93$&$10(6)$ & $92$&$26(7)$ & $0$&$3955(6)$ & $0$&$3994(8)$ &
$0$&$79477(6)$ \\

POVMC     & 70  & 76  & $-34$&$88973(4)$  & $-34$&$09472(4)$  &
$93$&$23(6)$ & $92$&$33(7)$ & $0$&$4024(7)$ & $0$&$391(1)$ &
$0$&$79501(6)$  \\

PBVMC     & 275 & 291 & $-34$&$90509(6)$  & $-34$&$1059(2)$   &
$98$&$03(6)$ & $96$&$8(1)$ & $0$&$1748(7)$ & $0$&$226(2)$ &
$0$&$7992(2)$   \\

POBVMC    & 292 & 300 & $-34$&$90475(6)$  & $-34$&$1055(2)$   &
$97$&$92(6)$ & $96$&$6(1)$ & $0$&$1483(7)$ & $0$&$222(2)$ &
$0$&$7993(2)$   \\

PDMC      & --  & --  & $-34$&$9026(2)$   & $-34$&$1072(2)$   &
$97$&$25(9)$ & $97$&$3(1)$ & \multicolumn{2}{c}{--} &
\multicolumn{2}{c}{--}         & $0$&$7954(3)$   \\

PODMC     & --  & --  & $-34$&$9028(2)$   & $-34$&$1074(2)$   &
$97$&$31(9)$ & $97$&$3(1)$ & \multicolumn{2}{c}{--} &
\multicolumn{2}{c}{--}         & $0$&$7954(3)$   \\

PBDMC     & --  & --  & $-34$&$9114(2)$   & $-34$&$1141(2)$   &
$100$&$00(9)$ & $100$&$0(1)$ & \multicolumn{2}{c}{--} &
\multicolumn{2}{c}{--}         & $0$&$7973(3)$   \\

POBDMC    & --  & --  & $-34$&$9099(2)$  & $-34$&$1128(2)$    &
$99$&$53(9)$ & $99$&$5(1)$ & \multicolumn{2}{c}{--} &
\multicolumn{2}{c}{--}         & $0$&$7971(3)$   \\

\hline \hline
\end{tabular}

\end{footnotesize}

\vspace{1em}

\end{center}

\noindent Table 1: Numbers of parameters in trial wave functions,
ground-state energies, percentages of correlation energies retrieved,
energy variances, and IP's of all-electron Ne (and Ne$^+$) and
pseudo-Ne (and pseudo-Ne$^+$), obtained using various methods.

\newpage

\textbf{Figure captions}

\begin{enumerate}
\item DMC energy of the all-electron Ne atom and the pseudo-Ne atom as
a function of the time step. The pseudo-Ne results are offset so that
the zero-time-step energy matches the all-electron energy. The
statistical error bars are smaller than the symbols. ``GF mods''
refers to the modifications to the DMC Green's function proposed in
Ref.~\cite{umrigar_1993a}.  The ``exact'' nonrelativistic infinite
nuclear mass result is taken from
Ref.~\cite{chakravorty_1993}. \label{fig:ne_tstep_bias}
\item The same as Fig.~\ref{fig:ne_tstep_bias}, but for a Ne$^+$
  ion. \label{fig:ne_plus_tstep_bias}
\item IP of all-electron and pseudo Ne as a function of the time
step. ``GF mods'' refers to the modifications to the DMC Green's
function proposed in Ref.~\cite{umrigar_1993a}. The experimental
result is taken from Ref.~\cite{kaufman_1972}, and the exact
nonrelativistic IP is from Ref.~\cite{chakravorty_1993}.  The
statistical error bars are smaller than the symbols.
\label{fig:ne_ion_tstep_bias}
\end{enumerate}

\newpage

\textbf{Figure \ref{fig:ne_tstep_bias}}

\vspace{3cm}

\begin{center}
\includegraphics{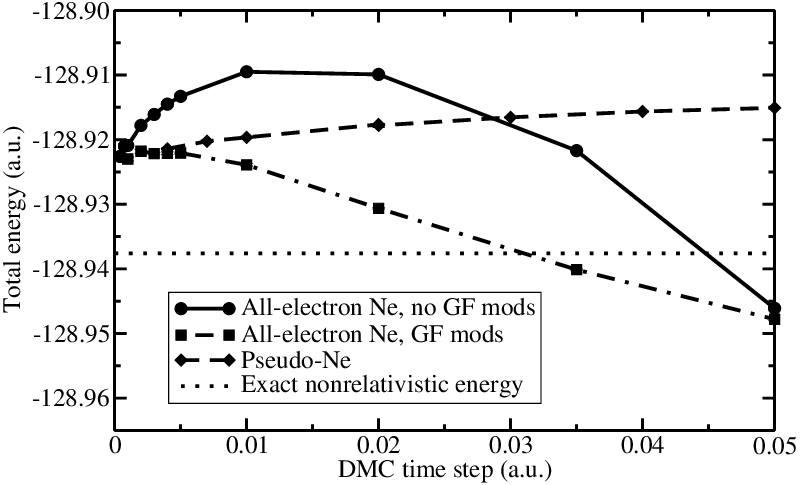}
\end{center}

\newpage

\textbf{Figure \ref{fig:ne_plus_tstep_bias}}

\vspace{3cm}

\begin{center}
\includegraphics{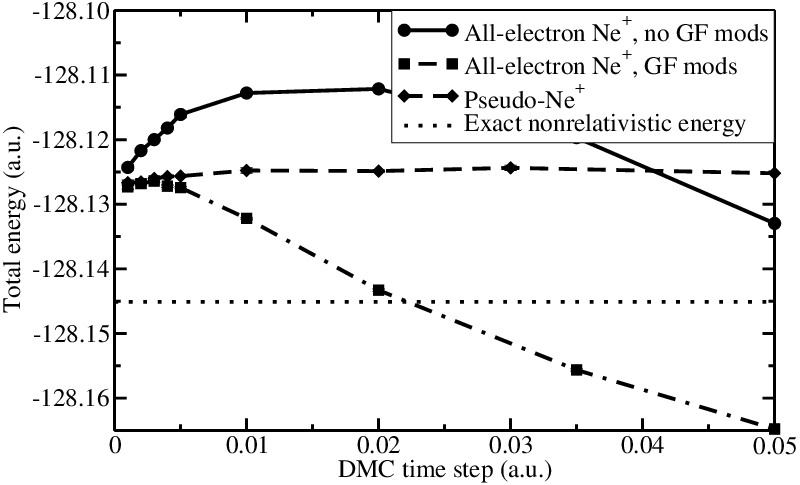}
\end{center}

\newpage

\textbf{Figure \ref{fig:ne_ion_tstep_bias}}

\vspace{3cm}

\begin{center}
\includegraphics{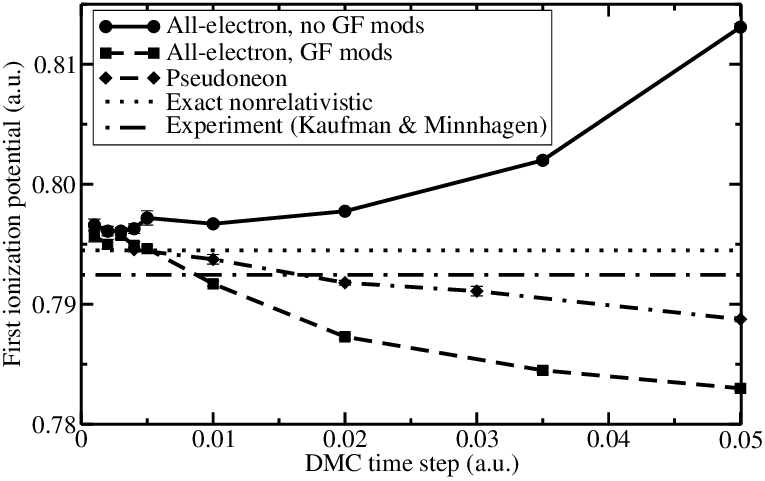}
\end{center}

\end{document}